\pdfoutput=1

\documentclass[11pt]{article}

\usepackage{acl}

\usepackage{times}
\usepackage{latexsym}
\usepackage{algorithm}
\usepackage{algpseudocode}
\usepackage{multirow}
\usepackage{colortbl}

\usepackage[T1]{fontenc}

\usepackage[utf8]{inputenc}

\usepackage{microtype}

\usepackage{inconsolata}

\usepackage{graphicx}

\usepackage{booktabs}
\usepackage{amsmath}

\title{Challenges in Trustworthy Human Evaluation of Chatbots}

\author{
  Wenting Zhao \quad Alexander M. Rush \quad Tanya Goyal\\
  \texttt{\{wz346,tg436\}@cornell.edu} \\}

\begin{document}
\maketitle
\begin{abstract}
Open community-driven platforms like Chatbot Arena that collect user preference data from site visitors have gained a reputation as one of the most trustworthy publicly available benchmarks for LLM performance. While now standard, it is tricky to implement effective guardrails to collect high-quality annotations from humans. In this paper, we demonstrate that three sources of bad annotations, both malicious and otherwise, can corrupt the reliability of open leaderboard rankings. In particular, we show that only 10\% of poor quality votes by apathetic (site visitors not appropriately incentivized to give correct votes) or adversarial (bad actors seeking to inflate the ranking of a target model) annotators can change the rankings of models by up to 5 places on the leaderboard. Finally, we discuss open challenges in ensuring high-quality human annotations.

\end{abstract}

\section{Introduction}
Reliable evaluation of free-form text generation quality is a long-standing challenge in NLP \cite{gehrmann2023repairing,celikyilmaz2020evaluation,goyal2022news}. 
Despite limitations, human annotation is widely accepted as the gold standard, especially for open-ended text generation tasks without an objective notion of correctness. As a result, platforms such as Chatbot Arena \cite{zheng2023judging,chiangchatbot} and WildVision Arena \cite{lu2024wildvision} that allow users to interact with available large language models (LLMs) and submit preference judgments for model pairs, have become extremely valuable resource in the NLP evaluation landscape. By providing free and easy access to available LLMs, these community-driven platforms are able to incentivize millions of user interactions\footnote{As of October 6, 2024, Chatbot Arena has collected 2,011,939 pairwise preference judgments.} and collect a large-scale and diverse dataset of user queries and preferences.  Deservedly, these peer production and community-driven platforms have emerged as one of the most trusted benchmarks in NLP today.\footnote{As an example, Google's Chief Scientist used high performance on Chatbot Arena to declare the success of their recent model release: \url{https://tinyurl.com/55xs2pz4}.}



Moreover, such benchmarks play a crucial role in auditing automatic evaluators by providing the necessary ground truth rankings that any evaluator can be validated against. In fact, the most popular automatic evaluation benchmarks today, including AlpacaEval~\cite{alpaca_eval}, WildBench \cite{lin2024wildbench}, MixEval \cite{ni2024mixeval} and Arena-Hard \cite{arenahard2024},  validate their metric by reporting high correlation with Chatbot Arena judgments.

Given its far-reaching impact, both on human and automatic benchmarking of LLMs, and consequently on LLM research more broadly, it is crucial to ensure that the model rankings on these open community leaderboards are trustworthy. However, challenges with obtaining high-quality human judgments from non-expert crowdworkers like Chatbot Arena users are widely discussed in literature \cite{karpinska2021perils, clark2021all,hoskinghuman}. Moreover, these platforms typically implement minimal quality controls for verifying annotation quality such as attention checks, user verification, etc. This sits in direct opposition to the goals of trustworthiness. 
In this paper, we play devil's advocate and ask: \textbf{is it even possible to ensure the reliability of a community-driven open platform, like Chatbot Arena, without sacrificing user scale?}

\begin{figure}
    \centering
    \includegraphics[scale=0.36,trim=50mm 220mm 0mm 50mm]{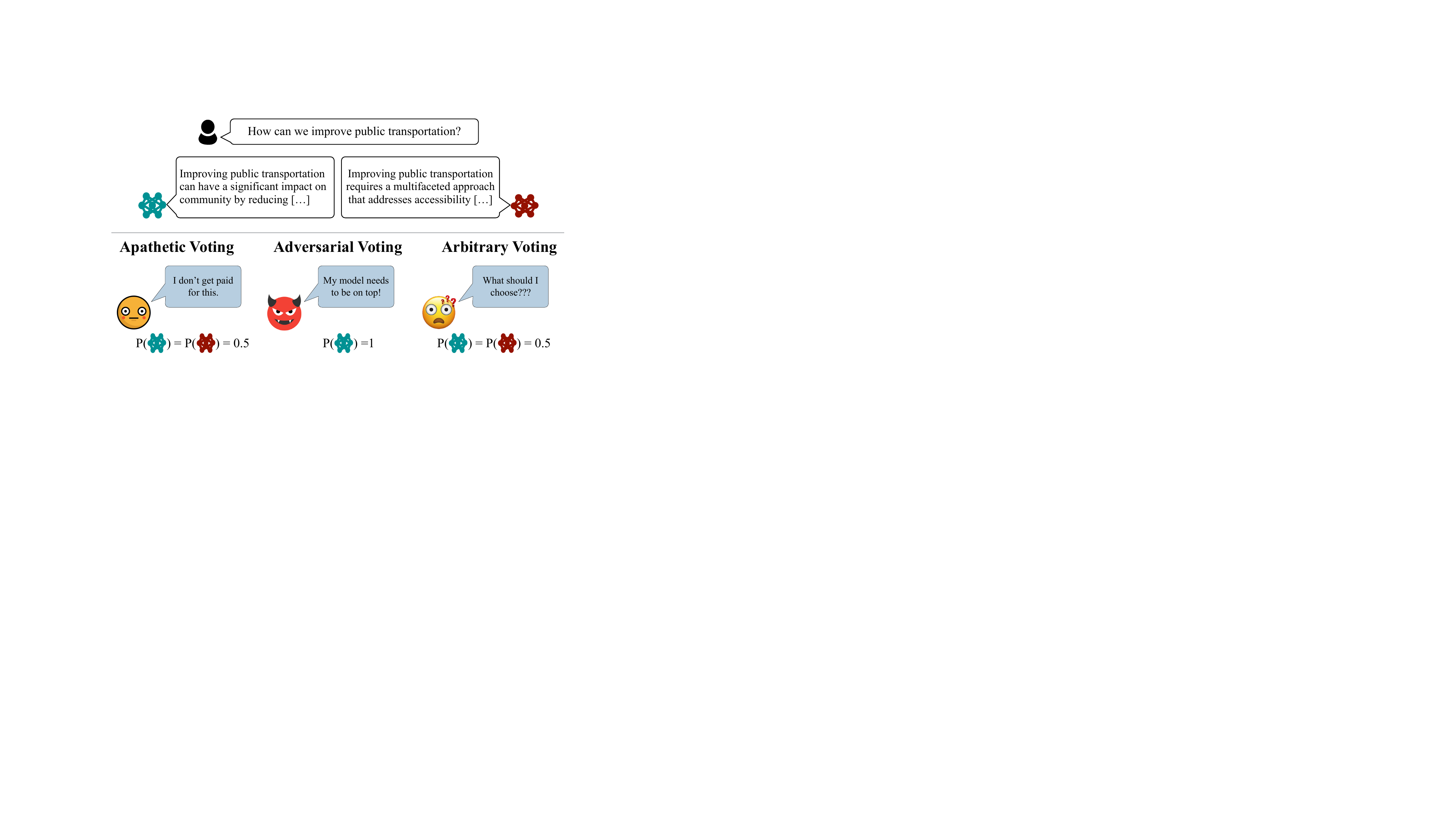}
    \caption{Our characterization of sources of poor-quality votes on open data annotation platforms: (1) Apathetic: Users who lack intrinsic motivation may submit random votes. (2) Adversarial: Malicious users aim to manipulate rankings by upvoting a target model. (3) Arbitrary: Users voting based on subjective preferences in response to open-ended questions.}
    \label{fig:overview}
\end{figure}

We approach this thought experiment from two angles. First, using Chatbot Arena as a case study, we consider three different sources of poor quality preference judgments or votes in the collected dataset: un-incentivized or \textbf{apathetic} users providing random judgments (Section~\ref{sec:apathetic}), malicious actors launching \textbf{adversarial} attacks to detect and artificially inflate a target model's ranking (Section~\ref{sec:adversarial}), and the inherent \textbf{arbitrariness} of preference votes for open-ended and subjective queries (Section~\ref{sec:arbitrary}). For the former two sources of votes, we show that small fractions of poor-quality judgments (either apathetic or adversarial) can have a non-trivial impact on the target models' rankings. 
Concerningly, poor annotations from either apathetic or adversarial voting are not easy to detect in a post-hoc manner. Moreover, even carefully recruited and onboarded human annotators exhibit low inter-annotator agreement on subjective queries, making inter-annotator-based techniques to filter out low-quality annotations ineffective.

Finally, we discuss open challenges in ensuring the reliability and human annotation quality in open-source community-driven benchmarks (Section~\ref{sec:discussion}).  We strongly believe that open data collection platforms offer an invaluable resource for the academic community and have facilitated essential work in developing new automatic evaluators \cite{alpaca_eval,lin2024wildbench,ni2024mixeval}, training and evaluating reward models \cite{lambert2024rewardbench}, etc. However, critical questions exist about their reliability, especially against adversarial attacks. We hope that our work will spur future research on quality control mechanisms for open platforms that power LLM evaluations.







\section{Background}
\label{sec:background}
In this paper, we run experiments with Chatbot Arena \cite{zheng2023judging,chiang2024chatbot} as a case study, although our insights are broadly applicable to other similar community-driven preference collection platforms. Below, we describe the preference collection pipeline and quality control measures employed by Chatbot Arena. 

\paragraph{Notation} Assume there are $k$ different models $\mathcal{M} = \{m_1, m_2, ..., m_k\}$ that need to be ranked on the leaderboard. Each new user on the platform submits a query $x$ and receives outputs from two different models $y_i \sim m_i(x)$ and $y_j \sim m_j(x)$.\footnote{The arena employs an adaptive sampling strategy that favors model pairs with higher uncertainty in relative performance, and also newly introduced models. However, exact details are not publicly shared, possibly to mitigate gaming.} The user has the option to submit a preference label $l \in \{i, j, \mathrm{tie}\}$. In order to ensure that this annotation is unbiased, the names of the models that the two outputs are sampled from is only revealed to end users after they have submitted their preference annotation. This arena logs data points of the form: $(x, y_i, y_j, m_i, m_j, l)$. 

These preferences are then used to estimate the pairwise win matrix between model pairs, i.e. $p(m_i>m_j)$. Next, they estimate the coefficients of the Bradley-Terry model \cite{bradley1952rank} to obtain scores $s_i$ for each model $m_i \in \mathcal{M}$. Models are sorted by $s_i$ to obtain the final ranking.


\paragraph{Quality control measures} The arena employs a list of filtering strategies: detecting malicious users according to a certain distribution (Section 5.1; \citet{chiang2024chatbot}), bot detection by Cloudflare and Google reCAPTCHA v3, automatic categorization pipelines to filter out low-quality data\footnote{\url{https://blog.lmarena.ai/blog/2024/hard-prompts/}}\footnote{\url{https://blog.lmarena.ai/blog/2024/arena-category/}}, placing limits on the number of votes each IP can provide in a day, and deduplicating top 0.1\% occurring prompts. However, these filtering strategies focus more on filtering bots than differentiating user votes with varying qualities. Therefore, we present results and discussions in this paper assuming minimal quality control checks in the backend to filter out bad quality user annotations\footnote{https://github.com/lm-sys/FastChat/}.

\paragraph{Released Artifacts} We conduct our experiments using the largest publicly released dataset by Chatbot Arena. It consists of 55k preference annotations\footnote{\url{https://huggingface.co/datasets/lmsys/lmsys-arena-human-preference-55k}}; it includes response pairs sampled from two of 64 unique models and the corresponding pairwise preference annotation. 

\section{Case Studies: Sources of Poor Quality Votes and Their Impact}
\label{sec:votes}

For our thought experiment, we hypothesize that there exist three potential sources of poor quality votes on open platforms: (a) apathetic votes by users that are un-incentivized, (b) adversarial votes that aim to inflate the ranking of a target model, and (c) arbitrary votes on difficult to meaningfully distinguish response pairs. For each of these, we study their impact on model rankings and the challenges in mitigating them.  


\subsection{Apathetic Voting}
\label{sec:apathetic}
The main attraction of open community platforms for end users is that they expose a free and easy-to-use API endpoint for LLMs. This incentivizes diverse users to interact with the platform and submit queries to explore their use cases. However, these platforms do not explicitly incentivize high-quality preference annotation. We hypothesize that at least r\% of users on the arena are apathetic and provide random or low-quality votes on the platform. 

\paragraph{Setup} We run experiments on Chatbot Arena's dataset of 55k preferences (discussed in Section~\ref{sec:background}). We assume that this dataset reflects ``true'' rankings of models based on gold human preferences. We study the change in model rankings for 3 arbitrarily selected models: Llama-2-7b-chat, Mistral-7b-instruct-v0.2, and Llama-2-13b-chat, assuming r\% of these preferences were instead assigned random labels by apathetic users during data collection.

\begin{table}[t]
    \centering
    \small
    \begin{tabular}{l|cccc}
    \toprule
        \textbf{Model} & \multicolumn{4}{|c}{\textbf{Leaderboard Ranking}} \\
        & Orig. & r=1 & r=5 & r=10 \\ \midrule
         Llama-2-7b-chat & 21 & 21 & 20\textsubscript{$\uparrow$1} & 21 \\
         Llama-2-13b-chat & 39 & 39 & 41\textsubscript{$\downarrow$2} & 34\textsubscript{$\uparrow$5} \\
         Mistral-7b-instruct-v0.2 & 36 & 38\textsubscript{$\downarrow$2} & 38\textsubscript{$\downarrow$2} & 41\textsubscript{$\downarrow$5} \\
    \bottomrule
    \end{tabular}
    \caption{Change in leaderboard rankings for 3 test models based on different percentages (r) of arbitrary votes. The subscripts denote gain ($\uparrow$) or loss ($\downarrow$) in rankings. We find that only 10\% poor quality annotations can change the rank of 2/3 systems by 5 places.}
    \label{tab:apathetic}
\end{table}

\paragraph{Results} Table~\ref{tab:apathetic} summarizes our results. \textbf{We find that only 10\% of apathetic votes in the dataset can change the leaderboard rankings of 2/3 models by 5 places} (namely Llama-2-13b-chat and Mistral-7b-instruct-v0.2).\footnote{Note that model frequency also impacts its susceptibility to ranking changes. All three models we inspect collectively occur in less than 10\% of all data samples.} Note that there are no existing studies characterizing the incentives or behaviors of an average user on open platforms like Chatbot Arena. Therefore, we have no way of estimating the fraction r of apathetic. 

\paragraph{Discussion: Can we detect and remove apathetic votes?} 
A major challenge in detecting apathetic votes is that they are often indistinguishable from arbitrary votes. Multiple past studies have found that output-level comparisons using a single label is ill-defined as an annotation task \cite{krishna2023longeval, goyal2022news} as users often rely on different criteria and disagree with each other. This ambiguity makes it hard to ascertain whether observed disagreements are due to personal variations in quality assessment (arbitrary voting, discussed further in Section~\ref{sec:arbitrary}) or due to apathetic or low-quality annotations by certain annotators. Despite challenges with detecting individual apathetic votes, detecting apathetic users may be viable by computing agreements between model rankings by individual users. This strategy is based on the intuition that while annotators might disagree on specific examples, their aggregate system-level judgments tend to be more aligned \cite{goyal2022news}. Finally, requesting additional justifications for votes, such as free-text rationales, can also help discourage apathetic votes. We discuss this more in Section~\ref{sec:discussion}.

\subsection{Adversarial Voting}
\label{sec:adversarial}
We assume there exists a malicious developer who seeks to inflate the rankings of their own target model $m_T$ on the arena leaderboard $A$. We argue that due to the lack of quality controls (e.g. user verification, attention checks, etc.), it is straightforward to inject preference votes for $m_T$ using a simple attack methodology.

Our main component is a \textbf{target model attribution algorithm} which, given a query-output pair $(q, y)$, predicts whether $y$ is sampled from the target model $m_T(q)$. 
Given such an algorithm, we can inflate the ranking of the target model $m_T$ using the following strategy: (1) Enter a prompt $q$ on the arena, (2) Detect if any of the two shown outputs $y_1$, $y_2$ are sampled from $m_T$, (3) If yes, vote for the target model $m_T$, (4) Repeat.


\paragraph{Target model attribution algorithm} We assume that the attribution algorithm has access to the target model logits. This is a reasonable assumption for our setting where a model developer seeks to inflate rankings. Our simple attribution algorithm is outlined in Algorithm 1 in Appendix~\ref{app:algo}.  

Essentially, we use teacher-forcing to determine the probability distribution over the vocabulary for all tokens at time step $t$, i.e. $P_{m_T}(.|x, y_1, ... y_{t-1})$. We sort the tokens in descending order of probability to identify the smallest subset of tokens that cover a cumulative probability mass of at least $p$. We compute the fraction of generation time steps $t$ for which the actual generated token $y_t$ falls within this top-p probability subset. We compare this against a threshold $t$ to classify generations $y = y_1 ... y_N$ as being sampled from $m_T$ or not. 

\addtolength{\tabcolsep}{-0.25em}
\begin{table}[t]
    \centering
    \small
    \begin{tabular}{l|ccccc}
    \toprule
        \textbf{Model} & \multicolumn{5}{|c}{\textbf{Leaderboard Ranking}} \\
        & Orig. & r=1 & r=5 & r=10 & r=100 \\ \midrule
         Llama-2-7b-chat & 21 & \cellcolor{red!25}23\textsubscript{$\downarrow$2} & 21 & \cellcolor{green!25}17\textsubscript{$\uparrow$4} & \cellcolor{green!25}1\textsubscript{$\uparrow$21} \\
         Llama-2-13b-chat & 39 & \cellcolor{green!25}36\textsubscript{$\uparrow$3} & \cellcolor{green!25}32\textsubscript{$\uparrow$5} & \cellcolor{green!25}28\textsubscript{$\uparrow$9} & \cellcolor{green!25}1\textsubscript{$\uparrow$39} \\
         Mistral-7b-instruct-v0.2 & 36 & \cellcolor{green!25}34\textsubscript{$\uparrow$2} & \cellcolor{green!25}34\textsubscript{$\uparrow$2} & \cellcolor{green!25}29\textsubscript{$\uparrow$7} & \cellcolor{green!25}2\textsubscript{$\uparrow$34} \\
    \bottomrule
    \end{tabular}
    \caption{Change in leaderboard rankings for 3 test models based on different percentages (r) of adversarial votes (upvoting the target model). We find that only 10\% adversarial annotations can change the rank of all systems by more than 4 places.}
    \label{tab:adversarial}
\end{table}

\begin{table}[t]
    \centering
    \small
    \begin{tabular}{c|ccc}
    \toprule
        \textbf{Model} & \textbf{TPR} & \textbf{TNR} & \textbf{\#Tokens} \\ \midrule
        Llama-2-7b-chat & 91.13 & 88.46 & 328.06 \\
        Llama-2-13b-chat & 100.00 & 89.93 & 326.53\\
        Mistral-7b-instruct-v0.2 & 91.28 & 86.69 & 319.46 \\
    \bottomrule
    \end{tabular}
    \caption{Intrinsic eval. of model attribution algorithm}
    \label{tab:intrinsic} 
\end{table}

\paragraph{Intrinsic Evaluation of Detector Algorithm} For all three test models, we report the true positive rate (TPR) and true negative rate (TNR) on the arena dataset in Table \ref{tab:intrinsic}. We find that our detector algorithm reports very high performance (e.g. TPR=91.13\%,  and TNR=88.46\% for Llama-2-7b-chat). We also find a positive correlation between the number of tokens and TPRs, which can be leveraged in the attack. 
Note that malicious actors can always improve the detector accuracy further using watermarking techniques \cite{kirchenbauer2023watermark}. Next, we use these highly performant models to cast adversarial votes.




\paragraph{Can we influence voting on the live Chatbot Arena platform?} We also implement a proof-of-concept of a real ``attack'' on Chatbot Arena to demonstrate that current guardrails, such as bot detection, can be bypassed easily. On October 13, 2024, we programmatically launched 100 queries into Chatbot Arena, extracted the two model responses, and successfully submitted a preference vote. To avoid contaminating the dataset, we only cast ``tie'' votes but note that it would be trivial to instead use the vote from the attribution algorithm. 

Interestingly, post-hoc analysis of this data revealed that \texttt{yi-lightning} family of models, released just 2 days later, were the most common (20\% of the responses) in this set.\footnote{Evenly distributed between \texttt{yi-lightning} and \texttt{yi-lightning-lite}.} We assume that Chatbot Arena had early access to these models and sampled them more frequently than others in order to collect enough votes. However, this knowledge of when particular models will be up-sampled can be easily exploited by adversaries to log a large fraction of upvotes for their model.

\paragraph{Impact of adversarial voting on leaderboard rankings} Similar to Section~\ref{sec:apathetic}, we run experiments on the 55k preference dataset from Chatbot Arena, assumed to reflect "true" votes. For 3 target models, we report the change in leaderboard rankings if adversarial voting was conducted on r\% of the data samples during data collection. Table~\ref{tab:adversarial} summarizes our results. Across all models, we show that adversarial attacks can substantially change leaderboard rankings if adversaries get to contribute 10\% votes for their model.\footnote{We assume that adversaries can get 10\% votes towards their own model because newly released models will be sampled more frequently.}. Note that, in this work, we only report results using the most simplistic version of this attack. We can further boost these numbers by not only upvoting the target model but also downvoting open-source competitor models or those ranked higher than the target model in the leaderboard.

\paragraph{Discussion: Can we detect and remove adversarial votes?} Open platforms can employ two types of mitigation strategies to address this issue: recognizing bot-like behavior to prevent votes from being cast, or detecting abnormal users post-hoc to filter out their votes. Platforms like Chatbot Arena already implement measures from both categories. For example, Chatbot Arena uses Cloudflare and Google reCAPTCHA to detect bots on their platform; however, we were able to bypass both programmatically. We did not find public information indicating that similar measures have been incorporated into the Wildvision Arena platform.

There are also opportunities to detect anomalous users post-hoc based on behaviors across multiple sessions or votes. Chatbot Arena implements a version of this strategy by comparing the distribution of ratings from a user (uniquely identified by IP address) against historical distributions to identify anomalies. Because committed adversaries may bypass these checks using IP rotation or similar techniques, we encourage further exploration of these approaches to make them more robust.

\subsection{Arbitrary Voting}
\label{sec:arbitrary}
We assume an idealized scenario where all users genuinely make their best effort to rank model outputs. However, we argue that holistically rating a response to an open-ended and inherently subjective query is ill-defined and liable to always be arbitrary. To demonstrate this, we conduct a small-scale annotation study for outputs of subjective \emph{Researchy} questions' prompts \cite{rosset2024researchy}.\footnote{Representative question: ``How can the education system be improved?''.}


\paragraph{Setup} We use these prompts and generate generate responses from four language models: Llama-3-8B, Llama-3-70B, GPT-4o, and GPT-3.5. We recruit four undergraduate CS students who are passionate about NLP and committed to providing thoughtful annotations. They evaluate responses on four dimensions: thesis, organization, reasoning, perspectives, and writing style. We offer them unlimited time and allow them to seek clarification from the authors when needed. Note that this dimension-wise rating is different from Chatbot Arena's setup of pairwise preferences. However, there already exist multiple prior works that argue that the task is under-defined in this latter setting and report low agreement between annotators \cite{goyal2022news,goyal2022snac,krishna2023longeval}. Therefore, we opt to run this study using a more well-defined task description.

\begin{table}[t]
\scriptsize
\centering
\begin{tabular}{lrrrrr}
\toprule
\textbf{} & \textbf{Th} & \textbf{Org} & \textbf{Re} & \textbf{Per} & \textbf{WS} \\ \midrule
GPT-3.5 vs GPT-4o & -0.36 & 5.51 & 9.91 & 17.18 & 20.06 \\
Llama-3-8b vs Llama-3-70b & 10.15 & -10.78 & 27.16 & 5.78 & 8.50 \\
Llama-3-8b vs GPT-3.5 & 11.34 & 7.19 & -12.15 & 3.53 & 7.45 \\
Llama-3-70b vs GPT-4o & 3.15 & -1.27 & 4.56 & 2.75 & -4.66 \\ \bottomrule
\end{tabular}
\caption{Fleiss' Kappa between four annotators on different evaluation axis: Th(esis), Org(anization), Re(asoning), Per(spectives), WS (Writing Style).}
\label{tab:kappa}
\end{table}

\paragraph{Results} Table \ref{tab:kappa} shows the inter-annotator agreement between the annotators. Overall, we find very low agreement between these well-intentioned annotators with clear guidelines, irrespective of the performance difference between the model pairs. More concerningly, the results highlight that traditional approaches like filtering out low-quality users/annotations using inter-annotator agreement may not be a viable strategy for open-ended queries as it is difficult to disentangle between of low inter-annotator agreement due to bad annotation (apathetic votes) or inherent subjectivity. Adversarial users can also ``hide'' their votes from similar scrutiny by using open-ended prompts for which vote choice is expected to be ambiguous.

\paragraph{Discussion} We argue that arbitrary votes are not ``noise'' and provide useful signals about models' relative performance. If most frontier models perform similarly well on a substantial fraction of real-world queries, this information should not be discarded but inform leaderboard Elo scores. Arbitrary votes become problematic when the majority of the leaderboard is dominated by open-ended queries that fail to meaningfully distinguish models, despite the existence of legitimate topics or skills along where models exhibit distinct behaviors. Identifying which test examples (or type of test examples) are most informative and up-weighting them when deriving aggregate scores are potential ways of addressing this \cite{rodriguez2021evaluation}.

\section{Conclusion \& Future Directions}
\label{sec:discussion}
Our experiments in Section~\ref{sec:votes} lay a convincing case for the need for stronger guardrails in open community-driven platforms. Although these are broadly accepted as the ground truth rankings of LLMs, we are concerned that it is easy to intentionally (adversarial) or unintentionally (apathetic, arbitrary settings) corrupt these leaderboards. The key challenge in mitigating the issue of poor quality annotations is: how can community-driven platforms strike the right balance between implementing necessary quality controls while also providing the right incentives and experience to users to continue to use these platforms. 

\paragraph{Richer feedback} 
We encourage the community to explore ideas from past research, such as soliciting fine-grained annotations \cite{krishna2023longeval, goyal2022snac}  or rationales \cite{mcdonnell2016relevant} in addition to the binary preference feedback. Rationales can be useful in encouraging apathetic users to think more critically about their votes (or abstain) and also for filtering out low-quality annotations from both apathetic and adversarial users. 

Past work in generation evaluation has discussed how binary preference, or even a single Likert rating, for the whole output, cannot meaningfully capture the nuances of human preferences \cite{gehrmanncase,gehrmann2023repairing}. Instead, fine-grained preference annotation is recommended, both along multiple dimensions or quality \cite{gehrmanncase} or for smaller units within the whole output \cite{krishna2023longeval,goyal2022snac}. More recent work proposes providing added context during evaluation to encourage higher agreement between annotators \cite{malaviya2024contextualized}. Future work must explore how these strategies can be incorporated into open platforms without inordinately increasing the annotation burden on users.

\paragraph{Stronger Guardrails}
Other guardrails could include reputation-based systems \cite{Adler2007ACR}, CAPTCHA \cite{von2003captcha,von2008recaptcha}, machine learning based anomaly detection \cite{kumar2014accurately,wu2016troll} and techniques that use annotator behavior traces on the platform to estimate quality \cite{goyal2018your}. 

\paragraph{Open access to collected dataset} Public release of the collected data on open platforms will spur research to address the annotation issues we discuss in this work. It would provide a more detailed overview into which types of queries are most well-equipped to distinguish between models, and what are the limitations of different families of models.



\section*{Limitations}
In this paper, we focus our analysis on one open community-driven platform, namely Chatbot Arena. However, there exist other similar platforms, like WildVision Bench, that implement similarly lax guardrails around annotation quality. Extending this analysis to such platforms can lead to added insights specific to vision language model evaluation. 

\section*{Acknowledgements}
We thank Deniz Bölöni-Turgut, Leo Lu, Aishanur Aydin, and Alicia Fulbright for providing model preference annotations on Researchy Questions. We thank the Chatbot Arena team for feedback on this draft. AMR was supported by NSF CAREER 2037519.

\bibliography{anthology,custom}

\appendix





\section{Model Attribution Algorithm} 
The model attribution algorithm used to carry out the adversarial attack in Section~\ref{sec:adversarial} is outlined below.
\label{app:algo}
\begin{algorithm}[ht]
\caption{Model Attribution}
\label{alg:model_attribution}
\begin{algorithmic}[1]
\Require Target model $m_T$, input sequence $x$, output sequence $y = (y_1, y_2, \dots, y_N)$, probability threshold $p$, decision threshold $t$
\Ensure 1 ($y$ is likely from $m_T$); 0 ($y$ is unlikely from $m_T$)
\State Initialize $\textit{results} \gets$ \textit{an empty list}
\For{$i = 1$ to $N$}
    \State $\mathcal{D}_i \gets \text{softmax}(P_{m_T}(x, y_1, \dots, y_{i-1}))$
    \State $S_i^*\gets \arg\min_{S_i} |S_i|$ s.t. $\sum_{t \in S_i} \mathcal{D}_i[t] \ge p$
    \If{$y_i \in S_i^*$}
        \State Append $1$ to $\textit{results}$
    \Else
        \State Append $0$ to $\textit{results}$
    \EndIf
\EndFor
\State Compute confidence score $c \gets \frac{\sum(\textit{results})}{N}$
\If{$c \ge t$}
    \State \Return 1
\Else
    \State \Return 0
\EndIf
\end{algorithmic}
\end{algorithm}

\end{document}